\date{\today}

\documentclass[aps,prb,twocolumn,superscriptaddress,groupedaddress]{revtex4}  
\usepackage{dcolumn,graphicx,amsmath,amssymb,algorithm,algpseudocode}
\usepackage{todonotes}
\usepackage{qcircuit}

\begin{document}

\definecolor{brickred}{rgb}{.72,0,0} 

\title{
Quantum Filter Diagonalization:
Quantum Eigendecomposition without Full Quantum Phase Estimation
}

\author{Robert M. Parrish}
\email{rob.parrish@qcware.com}
\affiliation{
QC Ware Corporation, Palo Alto, CA 94301
}

\author{Peter L. McMahon}
\affiliation{
QC Ware Corporation, Palo Alto, CA 94301
}
\affiliation{
School of Applied Engineering and Physics,
Cornell University, Ithaca, NY 14853
}

\begin{abstract} 
We develop a quantum filter diagonalization method (QFD) that lies somewhere
between the variational quantum eigensolver (VQE) and the phase estimation
algorithm (PEA) in terms of required quantum circuit resources and conceptual
simplicity. QFD uses a set of of time-propagated guess states as a variational
basis for approximate diagonalization of a sparse Pauli Hamiltonian. The
variational coefficients of the basis functions are determined by the
Rayleigh-Ritz procedure by classically solving a generalized eigenvalue problem
in the space of time-propagated guess states. The matrix elements of the
subspace Hamiltonian and subspace metric matrix are each determined in quantum
circuits by a one-ancilla extended swap test, i.e., statistical convergence of a
one-ancilla PEA circuit. These matrix elements can be determined by many
parallel quantum circuit evaluations, and the final Ritz estimates for the
eigenvectors can conceptually be prepared as a linear combination over separate
quantum state preparation circuits. The QFD method naturally provides for the
computation of ground-state, excited-state, and transition expectation values.
We numerically demonstrate the potential of the method by classical simulations
of the QFD algorithm for an $N=8$ octamer of BChl-a chromophores represented by
an 8-qubit \emph{ab initio} exciton model (AIEM) Hamiltonian. Using only a
handful of time-displacement points and a coarse, variational Trotter expansion
of the time propagation operators, the QFD method recovers an accurate
prediction of the absorption spectrum.
\end{abstract}


\maketitle

\section{Introduction}

The effective extraction of a few low-lying eigenpairs and corresponding
transition operator expectation values from a large Hermitian matrix specified
in sparse Pauli form is a ubiquitous task in mathematical physics and
optimization. In the former consideration, the operator to be diagonalized is
often a representation of the Hamiltonian for an interesting quantum system, and
the diagonalization amounts to solving the time-independent Schr\"odinger
equation within this representation.  In the latter consideration, the operator
to be diagonalized is a classical Ising Hamiltonian that is isomorphic to the
cost matrix of a binary optimization problem. Here, the matrix is already
diagonal in the $\hat Z$ basis, and the eigendecomposition amounts to an
exhaustive search for low-lying diagonal matrix elements.

Numerous classical approaches have been developed to approximately solve this
problem, but all require extensive validation efforts to ensure sufficient
accuracy for each new class of physical problem encountered. In particular, most
such approximations rely on perceived spatial-, rank-, or tensor-sparsity
structure in the physical problems at hand, in an attempt to ameliorate the
exponential naive classical cost of representing even a single eigenpair
directly. E.g., in the diagonalization of quantum Hamiltonians of electronic
structure theory, a myriad of approximate theories ranging from density
functional theory,\cite{
Hohenberg:1964:B864,
Kohn:1965:A1133,
Runge:1984:997,
Koch:2001:DFT}
coupled cluster theory,\cite{
Cizek:1966:4256,
Purvis:1982,
Crawford:2007:33,
Bartlett,
deustua2017converging,
deustua2018communication},
selected configuration interaction theory,\cite{
bender1969studies,
huron1973iterative,
booth2009fermion,
cleland2010communications,
holmes2016heat,
Schriber:2016:161106,
schriber2017adaptive}
and density matrix renormalization group theory\cite{
White:1992:2863,
Chan:2011:465}
has been developed - each performs remarkably well in certain regimes of the
problem space, and fails qualitatively in others.  A promising alternative
approach is to use a universal quantum computer to aid in part of the
computation. As increasingly powerful quantum circuit hardware becomes
available, this approach will remove formal tractability barriers regarding the
storage and manipulation of Hilbert space quantities, but will surely be fraught
with conceptual challenges involving the design of efficient hybrid
quantum/classical algorithms that will have only limited few-qubit quantum
circuit gates and that will necessarily output only quantum observable
measurements. Numerous impressive strides have been made over the past few
decades in developing efficient algorithms along these lines. A foundational
algorithm for quantum eigendecomposition is the phase estimation algorithm
(PEA).\cite{
kitaev1995quantum,
Abrams:1997:2586,
cleve1998quantum,
Abrams:1999:5162,
Aspuru:2005:1704,
Lanyon:2010:106,
Wecker:2014:022305,
Tubman:2018:X}
Unfortunately, phase estimation requires nested control of
Trotterized time propagation operations by a large array of ancilla qubits,
which, when expanded to the standard library of 1- and 2-qubit gates, yields
extremely long circuits which will not be tractable in the near term. Motivated
by the limited gate depths and low fidelity of extant noisy intermediate-scale
quantum (NISQ) devices,\cite{Preskill:2018:79} several compelling ``variational
quantum  algorithms'' have been developed over the past few years that feature
vastly lowered gate requirements (and usually no ancilla qubits) at the cost of
designing and optimizing a heuristic variational entangler circuit. An
archetypical method of this type is the variational quantum eigensolver
(VQE),\cite{
Peruzzo:2014:4213}
which has been widely deployed in simulators and in several types of quantum
hardware to target the lowest eigenstates of the Hamiltonian for electrons in
molecules and other physical systems.\cite{
Peruzzo:2014:4213,
McClean:2016:023023,
OMalley:2016:031007,
Kandala:2017:242,
McClean:2017:X,
Romero:2018:104008,
Nam:2019:IonWater,
GaoIBMBattery}
VQE has seen numerous recent extensions to efficiently treat excited
states,\cite{
Peruzzo:2014:4213,
McClean:2017:042308,
Higgott:2018:X,
Lee:2018:JCTC,
Colless:2018:011021,
Nakanishi:2018:VQE,
Parrish:2019:230401}
transition properties,\cite{
Parrish:2019:230401}
and gradient properties.\cite{
Obrien:2019:grad,
Mitarai:2019:grad,
Parrish:2019:grad}
A doppelg\"anger of VQE in the area of optimization is the quantum approximate
optimization algorithm (QAOA),\cite{
Farhi:2014:Quantum}
which also uses a specially constructed
variational entangler circuit to attempt to approximate diagonalization of a
classical Ising Hamiltonian. A large body of work has been devoted to the
practical utilization of QAOA and variants to provide direct optimization of
practically-relevant binary optimization problems.\cite{
Lucas:2014:Ising,  
farhi2017quantum,
otterbach2017unsupervised,
moll2018quantum,
crooks2018performance,
zhou2018quantum,
brandao2018fixed,
glover2018tutorial,
hadfield2019quantum,
gilyen2019optimizing,
nannicini2019performance}
 VQE and QAOA often produce accurate results with short quantum circuits
relative to more-traditional quantum algorithms, but the \emph{ad hoc}
definitions of the variational entangler circuits remain a significant barrier
to analysis and routine black-box deployment of these methods. Moreover, the
difficult nonlinear optimization procedure of the variational circuit parameters
has proven to be a difficult practical aspect of variational quantum algorithms,
with such conceptual nightmares encountered as ``barren plateaus'' of vast
regions of the parameter space that are far from optimal and have vanishing
gradient information.\cite{McClean:2018:4812}

In the present work, we develop an algorithm we refer to as ``quantum filter
diagonalization'' (QFD) that conceptually lives somewhere between VQE and PEA.
The algorithm starts from a set of easily classically prepared approximate
reference states for the targeted eigenvectors, and then builds a variational
ansatz of a linear combination of time-advanced and time-delayed reference
states. The preparation of time-advanced and time-delayed reference states are
conceptually prepared by application of the Trotterized time propagation
operator on quantum circuit hardware. Key to the algorithm is the evaluation of
both Hamiltonian and metric transition expectation values over different pairs
of time-advanced and time-delayed reference states: a single-ancilla variant of
the quantum swap test is found to be sufficient for this purpose. Finally, the
algorithm concludes with a classical diagonalization of the generalized
eigenvalue problem involving the quantum Hamiltonian and metric matrix elements
evaluated over time-advanced and time-delayed reference states. The resultant
classical eigenvalues are variational estimates of the eigenvalues of the full
Hamiltonian, while the classical eigenvectors provide a recipe for the classical
mixing of time-advanced and time-delayed reference states to reconstruct the
approximate eigenvectors in the full space. Post-facto evaluation of matrix
elements in the quantum circuit hardware then enables the extraction of
transition property expectation values over other sparse Pauli-basis Hermitian
operators. 

The use of a grid of time-propagated states as a basis for quantum
eigendecomposition algorithms is certainly not new - in fact, it can be argued
that the original phase estimation algorithm was founded on just such a basis.
Some time ago, Somma \emph{et al} proposed an
algorithm\cite{somma2002simulating} based on taking the discrete Fourier
transform of observables evaluated over a regularly-spaced time propagation
grid, but this method only works well when there are a small number of distinct
and well-separated eigenvalues. More recently, O'Brien \emph{et al} published an
interesting time-grid  method\cite{obrien2019quantum} based on a Prony-type
fitting of the relative time-shift overlap matrix.  Kyriienko has also published
a highly compelling ``quantum inverse iteration''
method\cite{kyriienko2019quantum} that uses a time propagation grid with a
predetermined quadrature recipe to approximate the operator $(\hat H -
\lambda)^{-\alpha}$ where $\lambda$ is a guess for the target eigenvalue and
$\alpha \geq 1$. When this operator is applied to a guess eigenstate, the
component of the eigenvector nearest to $\lambda$ is significantly magnified,
improving the solution. Very recently, as we were finalizing the numerical
demonstrations for this manuscript, Somma published yet another interesting time
grid method\cite{somma2019quantum} that uses a Fourier series for a smooth
cutoff function to extinguish high-lying eigenvectors. The novelty of our
approach is primarily in the explicit use of the time-propagated states as a
variational basis ansatz for the full eigenstates. This necessitates the
evaluation of the variational basis subspace Hamiltonian and metric matrices
(evaluated in the proposed method with quantum circuits by using an extension of
the swap test) followed by a classical generalized eigensolution.  We also point
out that our ``quantum filter diagonalization'' approach is heavily inspired by
the classical filter diagonalization
approach,\cite{neuhauser1990bound,neuhauser1994circumventing,grozdanov1995recursion,mandelshtam1995spectral,wall1995extraction,mandelshtam1997low}
which uses a basis of either time-propagated reference states or a Chebyshev
expansion of $(\hat H - \lambda)^{-\alpha}$ acting on reference states. The
first of these variants closely resembles the flavor of quantum filter
diagonalization developed here, while the second of these variants lies closer
to an extended version of the Kyriienko quantum inverse iteration method.
Notably, in classical filter diagonalization approaches, the same final
classical generalized eigenproblem arises as we encounter below. The key
difference between classical and quantum filter diagonalization is that the
Hamiltonian and metric matrix elements must be approximated by sparsity
considerations or Monte Carlo integration in the classical approach, while we
instead use a quantum circuit to efficiently evaluate the matrix elements in the
quantum approach. It also worth pointing out that our QFD method was heavily
inspired by Suzuki et al's recent method for amplitude estimation without phase
estimation,\cite{suzuki2019amplitude} in which is was shown that the PEA
portions of a  Grover-type amplitude estimation algorithm could be largely
replaced by a larger set of quantum measurements performed over a variety of
quantum circuits.

\section{Theory}

\subsection{Problem Statement}

Consider a system of $N$ qubits $\{ A \}$. We are given a Hamiltonian operator
in sparse Pauli form, e.g.,
\begin{equation}
\label{eq:Hamiltonian}
\hat H
\equiv
\sum_{A}
\mathcal{Z}_{A}
\hat Z_{A}
+
\mathcal{X}_{A}
\hat X_{A}
\end{equation}
\[
+
\sum_{A>B}
\mathcal{ZZ}_{AB}
\hat Z_{A}
\otimes
\hat Z_{B}
+
\mathcal{ZX}_{AB}
\hat Z_{A}
\otimes
\hat X_{B}
\]
\[
+
\mathcal{XZ}_{AB}
\hat X_{A}
\otimes
\hat Z_{B}
+
\mathcal{XX}_{AB}
\hat X_{A}
\otimes
\hat X_{B}
+
\ldots
\]
Here the $\ldots$ indicates the possible presence of 3-and higher-body terms,
but we do assume that the total number of terms scales polynomially in $N$. As
written with only $\hat X$ and $\hat Z$ terms, the Hamiltonian is real symmetric -
this is the case for all chemical Hamiltonians which we will encounter in the
numerical test cases.  However, all details of the QFD approach presented below
would remain valid if the Hamiltonian was a more-general Hermitian operator with
an imaginary antisymmetric component, e.g., $\hat Y$-type terms.

The objective is to determine an efficient recipe to conceptually prepare the
lowest few eigenstates $\{ | \Psi^{\Theta} \rangle \}$ and evaluate the
corresponding eigenvalues $\{ E^{\Theta} \}$,
\begin{equation}
\hat H
| \Psi^{\Theta} \rangle
=
E^{\Theta}
| \Psi^{\Theta} \rangle
: 
\
\langle \Psi^{\Theta} | \Psi^{\Theta'} \rangle
=
\delta_{\Theta \Theta'}
\end{equation}
And then to evaluate some low-order (transition) operator expectation values,
\begin{equation}
O^{\Theta \Theta'}
\equiv
\langle \Psi^{\Theta} |
\hat O
| \Psi^{\Theta'} \rangle
\end{equation}
Where $\hat O$ is also a low-order Hermitian Pauli operator. Note that the
observables $\{ E^{\Theta} \}$ and $\{ O^{\Theta \Theta'} \}$ are all that we
ultimately require
- the eigenstates $\{ | \Psi^{\Theta} \rangle \}$ are only conceptually useful,
  and never need to be explicitly represented classically.

\subsection{Spectrum Normalization}

In general, the Hamiltonian will present with an arbitrary spectral range
$[E^{\mathrm{min}}, E^{\mathrm{max}}]$. To normalize the time propagation grid,
we must determine a scaling parameter $\kappa$ that brings the spectrum into the
interval $[0, 1]$, modulo a constant phase factor,
\begin{equation}
\label{eq:kappa}
\kappa
\equiv
E^{+}
-
E^{-}
\geq
E^{\mathrm{max}}
-
E^{\mathrm{min}}
+
\Delta
\end{equation}
Here $\Delta$ is a small overage to provide a gap between the lowest and highest
eigenvalues upon mapping to the periodic domain $[0, 1]$. Efficient classical
heuristic techniques such as the Gershgorin circle theorem can often be used to
provide good estimates of $E^{-}$ and $E^{+}$. 

\subsection{QFD Ansatz}

The variational ansatz for quantum filter diagonalization (QFD) is,
\begin{equation}
| \Psi^{\Theta} \rangle
\equiv
\sum_{\Xi k}
C_{\Xi k}^{\Theta}
e^{-i 2 \pi k \hat H / \kappa}
| \Phi_{\Xi} \rangle
\equiv
\sum_{\Xi k}
C_{\Xi k}^{\Theta}
| \Gamma_{\Xi k} \rangle
\end{equation}
Here $\{ | \Phi_{\Xi} \rangle \}$ are a handful of easily-prepared guess states
that are determined by classical preprocessing and that can be efficiently
prepared by simple quantum circuits. The ``time'' index $k$ covers integers
from $-k_{\mathrm{max}}$ to $+k_{\mathrm{max}}$. As $k_{\mathrm{max}}
\rightarrow \infty$, the variational completeness of ansatz increases. The
operator $\hat U_{k} \equiv e^{-i 2 \pi k \hat H / \kappa}$ is the time
propagation operator, which provides a systematic way to extend the basis. We
reserve the right to make minor definitional modifications to $U_{k}$ later in
this work, e.g., by approximating by a Trotterization procedure.

The only free parameters of the ansatz are the values of $C_{\Xi k}^{\Theta}$,
which are determined through the variational Rayleigh-Ritz procedure by
classically solving the subspace eigenvalue problem,
\[
\sum_{\Xi' k'}
\mathcal{H}_{\Xi k, \Xi' k'}
C_{\Xi' k'}^{\Theta}
=
\sum_{\Xi' k'}
\mathcal{S}_{\Xi k, \Xi' k'}
C_{\Xi' k'}^{\Theta}
E^{\Theta}
:
\]
\begin{equation}
\label{eq:subspace}
\sum_{\Xi k}
\sum_{\Xi' k'}
C_{\Xi k}^{*\Theta}
\mathcal{S}_{\Xi k, \Xi' k'}
C_{\Xi' k'}^{\Theta'}
=
\delta_{\Theta \Theta'}
\end{equation}
The all-important ``quantum'' matrix elements are the subspace Hamiltonian,
\[
\mathcal{H}_{\Xi k, \Xi' k'}
\equiv
\langle \Gamma_{\Xi k} |
\hat H
| \Gamma_{\Xi' k'} \rangle
\]
\begin{equation}
\label{eq:Hsub}
=
\langle \Phi_{\Xi} | 
e^{+i 2 \pi k \hat H / \kappa}
\hat H
e^{-i 2 \pi k' \hat H / \kappa}
| \Phi_{\Xi'} \rangle
\end{equation}
and the subspace metric,
\[
\mathcal{S}_{\Xi k, \Xi' k'}
\equiv
\langle \Gamma_{\Xi k} 
| \Gamma_{\Xi' k'} \rangle
\]
\begin{equation}
\label{eq:Ssub}
=
\langle \Phi_{\Xi} | 
e^{+i 2 \pi k \hat H / \kappa}
e^{-i 2 \pi k' \hat H / \kappa}
| \Phi_{\Xi'} \rangle
\end{equation}
Techniques developed below will enable the efficient evaluation of these matrix
elements in terms of extended swap test quantum circuits with a single ancilla
qubit. The evaluation of these matrix elements is trivially parallelizable
across multiple quantum circuit hardware instances. Note that the variational
parameters are all determined monolithically \emph{after} evaluation of all
quantum circuit matrix elements. Moreover, the classical generalized
eigenproblem encountered has been extensively studied in classical electronic
structure methods,\cite{SzaboAndOstlund} and can be solved with a pair of calls
to a standard eigenproblem solver such as \textsc{Lapack's} \texttt{ZHEEV}. A
spectral cutoff in the eigenvalues of the metric matrix can be used to
ameliorate poor condition encountered during determination of the inverse
square root of the metric matrix, a procedure referred to in electronic
structure methods as canonical orthogonalization.

\subsection{Transition Properties}

After the completion of the algorithm, transition property expectation values
over another Hermitian operator $\hat O$ in sparse Pauli form may be evaluated
as,
\begin{equation}
\label{eq:transition}
\langle \Psi^{\Theta} |
\hat O
| \Psi^{\Theta'} \rangle
=
\sum_{\Xi k}
\sum_{\Xi' k'}
C_{\Xi k}^{*\Theta}
\mathcal{O}_{\Xi k, \Xi' k'}
C_{\Xi' k'}^{\Theta'}
\end{equation}
where the quantum matrix elements of the property operator are,
\begin{equation}
\label{eq:Osub}
\mathcal{O}_{\Xi k, \Xi' k'}
\equiv
\langle \Gamma_{\Xi k} |
\hat O
| \Gamma_{\Xi' k'} \rangle
\end{equation}
\[
=
\langle \Phi_{\Xi} | 
e^{+i 2 \pi k \hat H / \kappa}
\hat O
e^{-i 2 \pi k' \hat H / \kappa}
| \Phi_{\Xi'} \rangle
\]

\subsection{General Quantum Matrix Elements}

Consider the $N$-qubit quantum states,
\begin{equation}
| A \rangle
\equiv
\hat V 
| \Omega \rangle
\end{equation}
\begin{equation}
| B \rangle
\equiv
\hat W
| \Omega\rangle
\end{equation}
Here, $| \Omega \rangle$ is an arbitrary reference state, and $\hat V$ and $\hat
W$ are arbitrary unitary operators. For an arbitrary Hermitian operator $\hat
O$, we wish to evaluate the transition expectation value (generally a complex
scalar quantity),
\begin{equation}
\langle A | \hat O |B \rangle
\equiv
\langle \Omega | \hat V^{\dagger} \hat O \hat W | \Omega \rangle
\end{equation}
This can be accomplished by the following $N+1$-qubit quantum circuit with a
single ancilla qubit,
\begin{equation}
\label{eq:swap2}
| \aleph \rangle
\equiv
\phantom{fuc}
\begin{array}{l}
\Qcircuit @R=0.1em @C=0.3em {
\lstick{|0\rangle}
 & \gate{H}
 & \ctrlo{1}
 & \ctrl{1}
 & \qw \\
\lstick{|\Omega\rangle}
 & \qw
 & \gate{V}
 & \gate{W}
 & \qw \\
}
\end{array}
=
\frac{1}{\sqrt{2}}
\left [
| 0 \rangle \otimes \hat V | \Omega \rangle
+
| 1 \rangle \otimes \hat W | \Omega \rangle
\right ]
\end{equation}
It can be easily shown that,
\begin{equation}
\langle A | \hat O |B \rangle
=
\langle \aleph | \hat X \otimes \hat O | \aleph \rangle
+
i
\langle \aleph | \hat Y \otimes \hat O | \aleph \rangle
\end{equation}
I.e., the required transition matrix elements between different basis states
over Hermitian Pauli operators can be statistically estimated by Pauli
measurements of an extended swap test circuit with one ancilla qubit. Viewed
another way, this circuit is a somewhat generalized one-ancilla phase estimation
algorithm (PEA) circuit. 

\subsection{Specific Quantum Matrix Elements}

If brevity of notation and code were of prime importance, we would define
$|\Omega\rangle \equiv | 0 \rangle$ and absorb the definition of the reference
state into the controlled unitaries $\hat V$ and $\hat W$. However, this would
be somewhat wasteful in requiring the ancilla to control both time propagation
and state preparation circuit elements. Instead, we adopt the convention that $|
\Omega \rangle$ is defined to be an appropriately selected reference state, and
the controlled unitaries $\hat V$ and $\hat W$ perform only time propagation
operations. For the diagonal blocks $\Xi = \Xi'$, this is particularly easy: we
define $|\Omega \rangle \equiv |\Phi_{\Xi} \rangle$, $\hat V \equiv e^{-i 2 \pi k
\hat H / \kappa}$ and $\hat W \equiv e^{-i 2 \pi k' \hat H / \kappa}$ to obtain
$\mathcal{H}_{\Xi k, \Xi' k'}$ and $\mathcal{S}_{\Xi k, \Xi' k'}$. For the
off-diagonal blocks $\Xi \neq \Xi'$, the $\hat V$ and $\hat W$ operators remain
unchanged, but the $|\Omega\rangle$ state is redefined as one of the
``interfering reference states,'' defined as,
\begin{equation}
\label{eq:off-diag}
| \Phi_{\Xi \Xi'}^{\pm, \mathrm{Re}} \rangle
\equiv
\frac{1}{\sqrt{2}}
\left [
| \Phi_{\Xi} \rangle
\pm
| \Phi_{\Xi'} \rangle
\right ]
\end{equation}
and,
\begin{equation}
| \Phi_{\Xi \Xi'}^{\pm, \mathrm{Im}} \rangle
\equiv
\frac{1}{\sqrt{2}}
\left [
| \Phi_{\Xi} \rangle
\pm
i
| \Phi_{\Xi'} \rangle
\right ]
\end{equation}
E.g.,
\[
2 \mathrm{Re} \left (
\mathcal{H}_{\Xi k, \Xi' k'}
\right )
\equiv
2 \mathrm{Re} \left (
\langle \Gamma_{\Xi k} |
\hat H
| \Gamma_{\Xi' k'} \rangle
\right )
\]
\[
=
\langle \Phi_{\Xi \Xi'}^{+} | 
e^{+i 2 \pi k \hat H / \kappa}
\hat H
e^{-i 2 \pi k' \hat H / \kappa}
| \Phi_{\Xi \Xi'}^{+} \rangle
\]
\begin{equation}
-
\langle \Phi_{\Xi \Xi'}^{-, \mathrm{Re}} | 
e^{+i 2 \pi k \hat H / \kappa}
\hat H
e^{-i 2 \pi k' \hat H / \kappa}
| \Phi_{\Xi \Xi'}^{-, \mathrm{Re}} \rangle
\end{equation}
and,
\[
- 2 \mathrm{Im} \left (
\mathcal{H}_{\Xi k, \Xi' k'}
\right )
\equiv
- 2 \mathrm{Im} \left (
\langle \Gamma_{\Xi k} |
\hat H
| \Gamma_{\Xi' k'} \rangle
\right )
\]
\[
=
\langle \Phi_{\Xi \Xi'}^{+, \mathrm{Re}} | 
e^{+i 2 \pi k \hat H / \kappa}
\hat H
e^{-i 2 \pi k' \hat H / \kappa}
| \Phi_{\Xi \Xi'}^{+, \mathrm{Re}} \rangle
\]
\begin{equation}
-
\langle \Phi_{\Xi \Xi'}^{-, \mathrm{Re}} | 
e^{+i 2 \pi k \hat H / \kappa}
\hat H
e^{-i 2 \pi k' \hat H / \kappa}
| \Phi_{\Xi \Xi'}^{-, \mathrm{Re}} \rangle
\end{equation}
and similarly for $\mathcal{S}_{\Xi k, \Xi' k'}$. To complete the formal definition of the
matrix elements, $\hat O \equiv \hat H$ for $\mathcal{H}_{\Xi k, \Xi' k'}$ and $\hat O
\equiv \hat I$ for $\mathcal{S}_{\Xi k, \Xi' k'}$.

\subsection{Trotterization (for Quantum Hamiltonian Diagonalization)}

At this point, one might ask why we did not instantly exploit the
property of time translational invariance in the evaluation of the Hamiltonian
and metric quantum matrix elements. Explicitly, the elements of each $\Xi, \Xi'$
subblock of these two matrices are Toeplitz as written, i.e., $\mathcal{H}_{\Xi
k, \Xi' k'} = \mathcal{H}_{\Xi, \Xi'}^{k' - k}$. Unfortunately, for quantum
Hamiltonians, we cannot \emph{quite} evaluate such matrix elements in standard
quantum circuit hardware. The reason for this is that the exponential over a
direct sum of noncommuting Pauli operators must be approximated by a procedure
such as Trotterization (also called Suzuki-Trotter splitting). Trotterization
necessarily induces moderate errors relative to the desired exact time
propagation operator. It is plausible that one could reduce these errors to a
negligible level with sufficiently many Trotter steps, in which case the
Trotterized quantum matrix elements would be statistically indistinguishable
from the exact quantum matrix elements. In such a case, the translational
invariance property could be exploited to reduce the number of required quantum
matrix elements to linear in $k_{\mathrm{max}}$.  Unfortunately, this approach
is likely to require too many Trotter steps to be tractable on NISQ-era
hardware. Moreover, if any nontrivial errors arise from Trotterization, such an
approach would lose the variational property, and the clean picture of a
wavefunction ansatz would be lost. Instead, we have elected to inject the
Trotterization into the definition of the basis functions,
\begin{equation}
\label{eq:time-trotter}
| \Gamma_{\Xi k} \rangle
\equiv
e^{-i 2 \pi k \hat H / \kappa}
| \Phi_{\Xi} \rangle
\leftarrow
\hat U_{\mathrm{Trotter}}
\left (
2 \pi k \hat H / \kappa
\right )
| \Phi_{\Xi} \rangle
\end{equation}
This breaks the Toeplitz property of the quantum matrix elements (necessitating
the evaluation of a quadratic number of matrix elements in $k_{\mathrm{max}}$),
but retains the clear and numerically robust picture of a variational
wavefunction ansatz. For instance, the exact solution is still obtained as
$k_{\mathrm{max}} \rightarrow \infty$ with our approach, which is not obtained
if the Trotterization is applied post facto within the Toeplitz quantum matrix
elements. Moreover, we believe that the parallel evaluation of a quadratic
number of low-depth quantum matrix elements is much more likely to be feasible
on NISQ-era hardware than the evaluation of a linear number of high-depth quantum
matrix elements.

\subsection{Explicit QFD Procedure}

The explicit steps of the QFD procedure are summarized as follows:
\begin{enumerate}
\item Classically compute the sparse Pauli matrix elements of the Hamiltonian
and other desired observable operators, e.g., $\{ \mathcal{Z}_{A} \}$ and
similar in Equation \ref{eq:Hamiltonian}.
\item Determine the spectral scaling parameter $\kappa$ of Equation
\ref{eq:kappa} by a classical preprocessing approach such as a heuristic
estimate of the Gershgorin circle theorem.
\item Classically determine the characteristics of the guess states $\{
|\Phi_{\Xi} \rangle \}$ and develop efficient quantum circuits to prepare these
states (and linear combinations of pairs thereof). For quantum Hamiltonian
diagonalization, a heuristic approximation like configuration interaction
singles (CIS) might be used. 
\item Use the extended swap test quantum circuits of Equation \ref{eq:swap2} and
variants thereof described around Equation \ref{eq:off-diag} to statistically
estimate the matrix elements of the subspace Hamiltonian $\mathcal{H}_{\Xi k,
\Xi' k'}$ of Equation \ref{eq:Hsub} and the subspace metric $\mathcal{S}_{\Xi k,
\Xi' k'}$ of Equation \ref{eq:Ssub}. Importantly, the Pauli expectation values
involved in both operators should be computed from the same set of quantum
circuit measurements, to maximize the propensity for error cancellation due to
correlated sampling. Trotterization of the time-propagation operators should be
included variationally as described in equation \ref{eq:time-trotter}.
\item Classically solve the subspace generalized eigenvalue problem of Equation
\ref{eq:subspace} to determine the QFD variational parameters $\{ C_{\Xi
k}^{\Theta} \}$ and the Ritz estimates of the eigenvalues $\{ E^{\Theta} \}$.
\item Evaluate desired transition expectation values $\{ O^{\Theta \Theta'} \}$
through the subspace expectation value formula of Equation \ref{eq:transition}.
Importantly, the subpace operator matrix elements $\mathcal{O}_{\Xi k, \Xi' k'}$
of Equation \ref{eq:Osub} often can reuse the quantum circuit measurements
performed to obtain the subspace Hamiltonian and metric matrices in in Step 4
above.
\end{enumerate}

\section{Results and Discussion}

\subsection{Demonstration System}

For a practical exploration of the QFD approach, we have considered an \emph{ab
initio} exciton model (AIEM) Hamiltonian\cite{
Sisto:2014:2857,
Sisto:2017:14924,
Li:2017:3493}
 for a linear stack of $N=8$ truncated BChl-a chromophore units, stacked in an
aligned geometry and then allowed to geometrically relax with
$\omega$PBE($\omega=0.3$)/6-31G*-D3. An \emph{ab initio} exciton model was
constructed for this system using a TDA-TD-DFT $\omega$PBE($\omega=0.3$)/6-31G*
treatment\cite{
Tawada:2004:8425,
Vydrov:2006:234109} 
of the monomers and unrelaxed nearest-neighbor dipole-dipole couplings. This
system requires an 8-qubit representation, and has a real 2-local Pauli
Hamiltonian. The system setup and Hamiltonian matrix elements are reported in
the main text and supplemental material of our MC-VQE paper.\cite{
Parrish:2019:230401}

\subsection{Spectral Normalization}

\begin{figure}[h!]
\begin{center}
\includegraphics[width=3.2in]{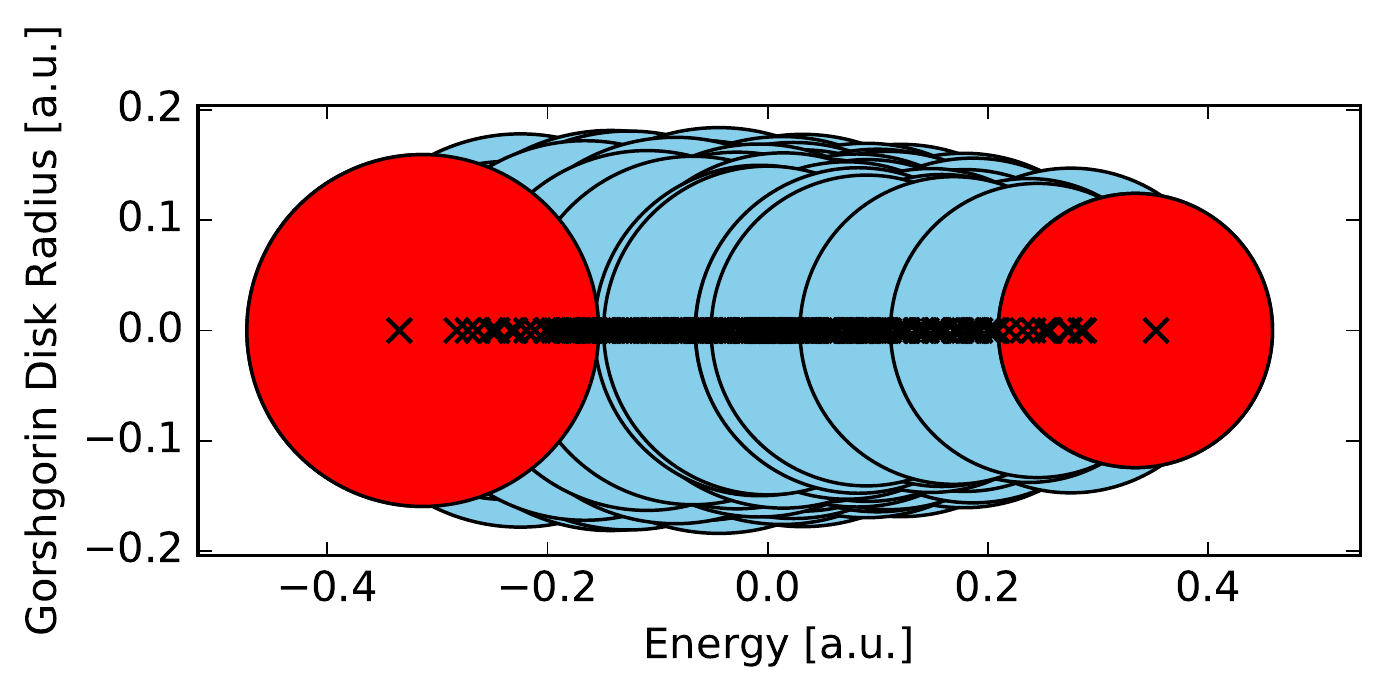}
\caption{Eigenspectrum and Gershgorin circle theorem analysis for $N=8$ BChl-a
AIEM Hamiltonian. The eigenvalues of $\hat H$ are given by small \texttt{x}
symbols on the x axis. The Gershgorin disks are presented as blue circles. The
Gershgorin disks for the first and last row of the Hamiltonian, corresponding to
the all-ground-state $|00\ldots \rangle$ configuration and all-excited-state
$|11\ldots \rangle$ configuration, are presented as red disks.}
\label{fig:disks-8}
\end{center}
\end{figure}

A prerequisite for QFD is a heuristic approach to estimate the spectral range of
$\hat H$ and determine the bounding parameter $\kappa$. Figure \ref{fig:disks-8}
depicts the full eigenspectrum and Gershgorin circle theorem analysis for this
system. The results are are congruent with our overall observations of
more-general sets of AIEM Hamiltonians: the spectral range generally grows
linearly with $N$, and the edges of the spectrum have significantly lower
density of states than the exponential crowding encountered in the middle of the
spectrum. The Gershgorin disks are generally wholly overlapping, i.e., there are
no topological gaps in the eigenspectrum. The Gershgorin disks for the first and
last row of the Hamiltonian, corresponding to the all-ground-state $|00\ldots
\rangle$ configuration and all-excited-state $|11\ldots \rangle$ configuration
are heuristically found to provide the extremal Gershgorin circles in all cases
we have encountered. This is not a rigorous result, but provides a good starting
point for more-advanced classical approaches that will rigorously bound the
spectrum in polynomial effort. Note that it is not generally possible to
explicitly classically evaluate all of the Gershgorin disks, as the size of the
diagonal of the Hamiltonian grows as $2^N$. For today, we will use these
heuristically extremal Gershgorin disks to determine $\kappa$. We note that this
actually provides a reasonably tight bound in practice: the true spectral range
is generally overestimated by only $30-50\%$ by the first/last Gershgorin disk
estimate for all AIEM systems we have yet encountered.

\subsection{Exact Time Propagation}

\begin{figure}[h!]
\begin{center}
\includegraphics[width=3.2in]{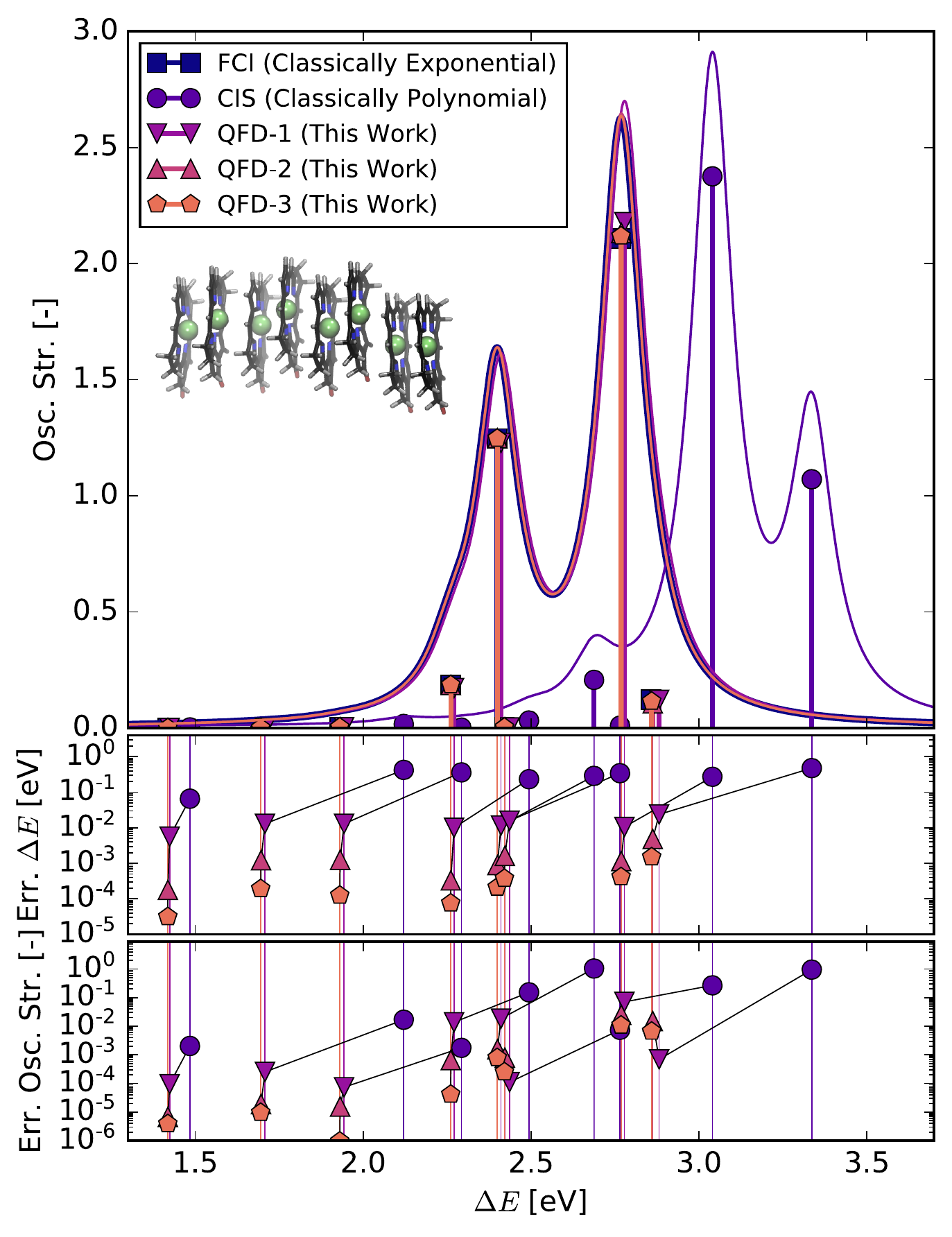}
\caption{
Test of simulated QFD with exact representation of the time propagation
operators vs. full configuration interaction (FCI) and configuration interaction
singles (CIS). The notation QFD-$k_{\mathrm{max}}$ means that the QFD time grid
is truncated at $k_{\mathrm{max}}$, e.g., QFD-1 has $k_{\mathrm{max}} = 1$ and
thus $k \in [-1, 0, +1]$. Top - Simulated absorption spectrum of $N=8$ linear
stack BChl-a test case (geometry depicted in inset), computed from the
excitation energies and oscillator strengths of the lowest 8 electronic
transitions, depicted as vertical sticks.  The envelope of the absorption
spectrum is sketched by broadening the contribution from each transition with a
Lorentzian with width of $\delta=0.15$ eV. Middle
- errors in excitation energies. Bottom - errors in oscillator strengths. Middle
  and bottom - thin lines are a guide for the eye. }
\label{fig:exact}
\end{center}
\end{figure}

Figure \ref{fig:exact} depicts a simulated absorption spectrum for the $N=8$
linear stack BChl-a test case. Excitation energies
$\Delta E^{\Theta} \equiv E^{\Theta} - E^{0}$ and oscillator strengths
$O^{0\Theta} \equiv (2/3) \Delta E^{\Theta} \langle | \Psi_{0} | \hat \mu |
\Psi_{\Theta} \rangle |^2$ are computed for each method, and compared to full
configuration interaction (FCI). Configuration interaction singles (CIS) including
the reference configuration $|00\ldots\rangle$ and all singly-excited configurations
$|10\ldots\rangle$, $|01\ldots\rangle$, \ldots are used to provide QFD guess states $\{ |
\Phi_{\Xi} \rangle \}$. In this example, 9 states are targeted, starting from 9
guess states. This example uses an exact representation of the time propagation
operator $\hat U_{k} \equiv e^{-i 2 \pi k \hat H / \kappa}$ representing an
ideal infinite-order Trotter expansion of the time propagation circuit. The
matrix elements $\mathcal{H}_{\Xi k, \Xi' k'}$ and $\mathcal{S}_{\Xi k, \Xi'
k'}$ are obtained by contraction of the relevant statevectors to model infinite
statistical sampling of the involved Pauli operators.  This test probes the
intrinsic limits of the QFD ansatz. The data indicates that the QFD series
converges rapidly for this test case - even using a single $k$ point, i.e.,
$k_{\mathrm{max}} = 1$, the improvement over the CIS guess is roughly one order
of magnitude for all states and for both excitation energies and oscillator
strengths. Moving to $k_{\mathrm{max}} = 2$, another order of magnitude
improvement is obtained in both properties, and the resultant spectrum is
visually indistinguishable from the reference full configuration interaction
(FCI) spectrum. Moving to $k_{\mathrm{max}} = 3$, additional improvement is
obtained, though the relative gains are somewhat smaller than in the first two
increments. By $k_{\mathrm{max}} = 3$, the errors in excitation energies are all
at or below $10^{-3}$ eV, while the errors in oscillators strengths are all at
or below $10^{-2}$ - essentially quantitative agreement. It is worth noting that
the errors in excitation energies and especially in oscillator strengths are
generally somewhat smaller and converge somewhat faster for lower-lying states. 

\subsection{Trotterized Time Propagation (Variational)}

\begin{figure}[h!]
\begin{center}
\includegraphics[width=3.2in]{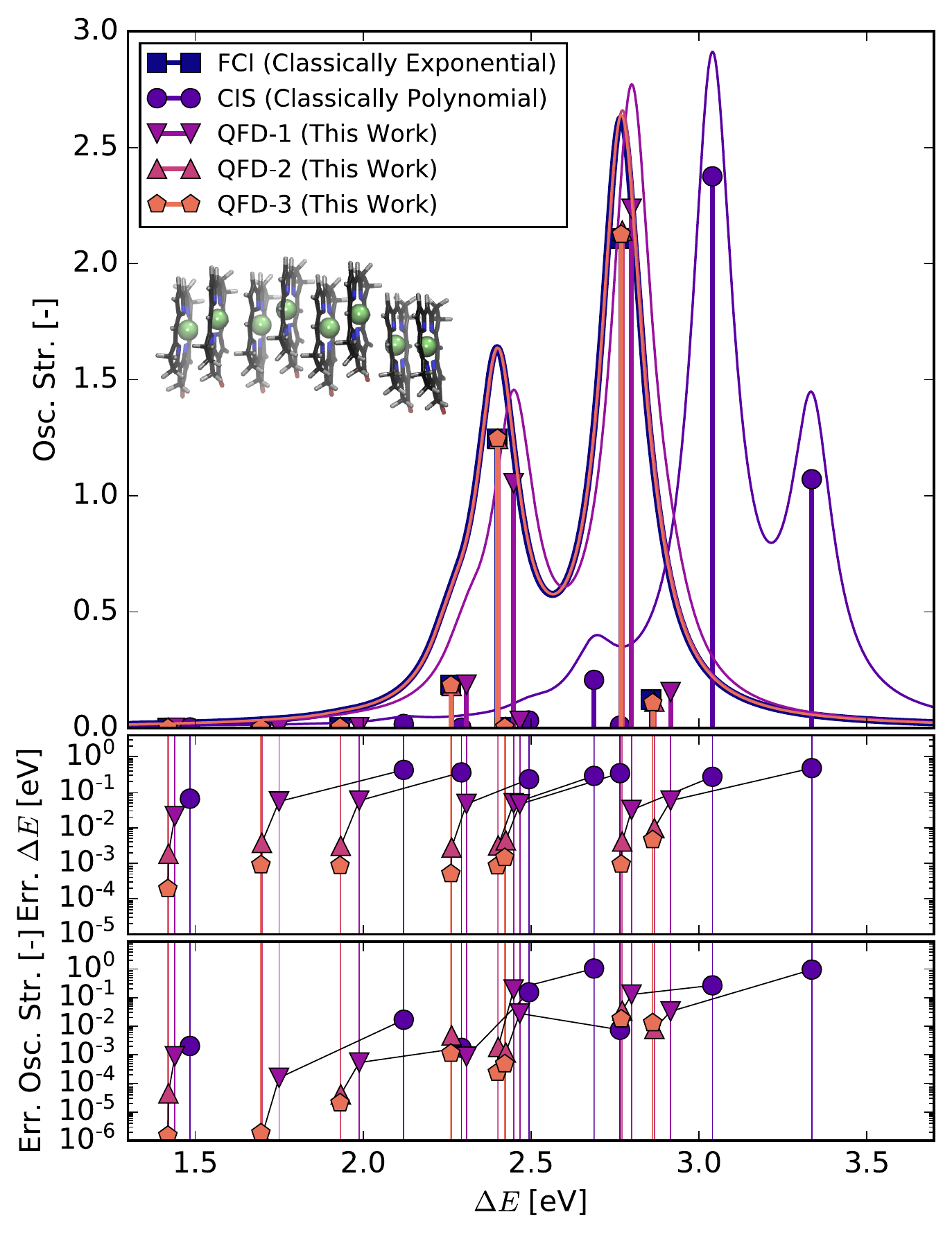}
\caption{
Test of simulated QFD with Trotterized representation of the time propagation
operators vs. full configuration interaction (FCI) and configuration interaction
singles (CIS). One Trotter step per $k$ point is used. The notation
QFD-$k_{\mathrm{max}}$ means that the QFD time grid is truncated at
$k_{\mathrm{max}}$, e.g., QFD-1 has $k_{\mathrm{max}} = 1$ and thus $k \in [-1,
0, +1]$. Top - Simulated absorption spectrum of $N=8$ linear stack BChl-a test
case (geometry depicted in inset), computed from the excitation energies and
oscillator strengths of the lowest 8 electronic transitions, depicted as
vertical sticks.  The envelope of the absorption spectrum is sketched by
broadening the contribution from each transition with a Lorentzian with width of
$\delta=0.15$ eV. Middle
- errors in excitation energies. Bottom - errors in oscillator strengths. Middle
  and bottom - thin lines are a guide for the eye. }
\label{fig:trotter}
\end{center}
\end{figure}

Figure \ref{fig:trotter} depicts the equivalent of Figure \ref{fig:exact}, but
with a physically realizable Trotter expansion of the time propagation
operators. Here a single first-order Trotter step is applied for each unit of
$k$ (e.g., 2 Trotter steps for $k=2$). The first-order Trotter step is of the
form,
\[
e^{-i 2 \pi k \hat H / \kappa}
\approx 
e^{-i 2 \pi k \hat H_{XX\ldots} / \kappa}
e^{-i 2 \pi k \hat H_{XZ\ldots} / \kappa}
\]
\begin{equation}
\times
e^{-i 2 \pi k \hat H_{ZZ\ldots} / \kappa}
e^{-i 2 \pi k \hat H_{ZX\ldots} / \kappa}
\end{equation}
The $\hat X_{A}$ one-body terms are grouped with $\hat H_{XX\ldots}$ and the
$\hat Z_{A}$ one-body terms are grouped with $\hat H_{ZZ\ldots}$. Comparing the
data between Figures \ref{fig:trotter} and \ref{fig:exact}, it is apparent that
even the remarkably coarse Trotter expansion employed here only marginally
degrades the accuracy of the QFD ansatz. QFD-1 with Trotterization produces a
more qualitatively divergent absorption spectrum than with exact time
propagation, but the Trotterized QFD-2 is essentially quantitatively converged.
Considering the lower panels of the figures, the Trotterized errors and
convergence rates are marginally slower than with ideal time propagation.
Overall, this is somewhat remarkable - the coarseness of the Trotterization
employed here means that the Trotterized vs. exact time propagation operators
differ significantly, but the QFD errors stemming from Trotterization at a given
$k_{\mathrm{max}}$ are generally of the same order of magnitude as changing from
to the previous $k_{\mathrm{max}}$ to the next. We attribute this tolerance of
coarse Trotterization to the use of a variational QFD ansatz: The Trotterized
QFD basis states are significantly different than the exact basis states, but the
coefficients are separately variationally optimized for each case, leading to
only marginal accuracy degradation with coarse Trotterization.

A non-variational form of Trotterization was also implemented, and found to give
exceedingly large errors even with small Trotter timesteps. This indicates that
the variational property of the QFD ansatz is critical to providing an accurate
representation, as expected. More details are provided in Appendix A below.

\section{Summary and Outlook}

We have discussed a quantum filter diagonalization method (QFD) with a set of
time-propagated guess states forming a variational basis for the approximate
Rayleigh-Ritz diagonalization of sparse Pauli operators. The variational
parameters of the method are determined through a one-shot classical solution of
a generalized eigenproblem in a small subspace of the full Hilbert space, while
observations of quantum circuits are used to compute the subspace matrix
elements needed as the input of this generalized eigenproblem. The method
converges monotonically toward the exact eigenpairs with a single discrete
$k_{\mathrm{max}}$ parameter determining the completeness of the time grid
expansion, and has been shown to be naturally applicable to the computation of
ground- and excited-state eigenvalues and transition properties.  Ideal
classical simulations of the method have been shown to give accurate results for
an example 8-qubit study involving the computation of the emph{ab initio}
exciton model (AIEM) absorption spectrum of a stack of 8 BChl-a chromophores.
High accuracy is obtained for both excitation energies and oscillator strengths
with only a handful of $k$ points and with remarkably coarse Trotter expansions
of the time propagation operators. The rather minor degradation of performance
upon Trotterization is attributed to the variational property of the Trotterized
QFD basis functions - a conceptual experiment involving a nonvariational variant
of QFD where Trotterization is performed after formal algebraic simplifications
of the quantum matrix elements yielded vanishingly accurate results. 

The QFD method occupies an interesting place relative to other quantum
algorithms for eigendecomposition. When compared to VQE, it appears that the QFD
controlled swap test circuits might be only marginally longer than VQE entangler
circuits, while the straightforward QFD ansatz may remove some of the conceptual
difficulties with designing an optimizing heuristic VQE entangler circuits.
Relative to PEA, QFD uses many more evaluations of short quantum circuits to
build toward a complete picture of the relevant subblock of the partially
diagonalized Hamiltonian. Therefore, QFD is potentially much more tractable with
NISQ-era quantum hardware than QFD, but will necessarily rely on spectral
structure such as nonvanishing average gaps in the relevant energy windows to
produce accurate results with short time expansions. Another interesting
distinction between PEA and QFD is that the former extracts the eigenvalues of
a unitary operator while the latter extracts the eigenvalues of a Hamiltonian
operator. In an ideal world, these would contain equivalent information which
would be extractable through knowledge of the exponential map. However, in a
Trotterized environment, QFD has the intriguing possibility that the quality of
the Hamiltonian eigenvalue estimate does not appear to be limited by the quality
of the Trotterization. 

Relative to other recent time-grid quantum algorithms, QFD appears to occupy a
somewhat different portion of the algorithmic landscape in the emergence,
explicit computation, and subsequent diagonalization of a variational subspace
eigenproblem. This avoids the requirement to parametrize an explicit filter
function with detailed knowledge of the approximate eigenspectrum, e.g., there
is no need to define a spectral cutoff in the QFD method.  To some extent, the
method resembles the Davidson approach of classical electronic structure theory
where one is able to compute and store full matrix-vector products, and
iteratively diagonalizes a subspace Hamiltonian. However, the Davidson procedure
has an additional advantage over QFD in that the Ritz estimates for the
eigenpairs from each stage of the decomposition are used to ``boost'' the
convergence by preconditioning.  This was recently demonstrated in the quantum
inverse iteration method for ground-state eigendecomposition, where the action
of $(\hat H - \lambda)^{-\alpha}$ on a guess state was expanded as a time grid,
and used to amplify the component of the eigenpair closest to $\lambda$ in the
guess state.  It would be very interesting to consider merging some of the
boosting ideas from quantum inverse iteration with the subspace diagonalization
of quantum filter diagonalization to obtain a method even closer in spirit to
the Davidson approach. One might also consider other approaches where a
classical solution of a subspace eigenproblem over quantum-enabled basis states
is invoked. For instance, one could imagine constructing a basis of
state-specific VQE states $| \Gamma_{\Xi} \rangle \equiv \hat U_{\Xi} |
\Phi_{\Xi} \rangle$, diagonalizing the subspace eigenproblem to determine the
mixing of the basis states, and then iteratively optimizing the VQE parameters
to minimize the sum of the resultant eigenvalues. Overall, all of these
approaches are moving toward an environment in which the target eigenvector is
approximated by a classical weighted sum of statevectors prepared by different
quantum circuits.

It will be useful to explore the characteristics of QFD in practice. One
challenge will be the design of physical quantum circuits performing the
one-ancilla controlled swap tests needed for the subspace matrix elements.
Another potential challenge is the accurate solution of the subspace
eigenproblem in the presence of statistical or device noise channels. Here, it
seems plausible that correlations in the noise between the subspace Hamiltonian
and subspace metric matrices (which may be evaluated simultaneously with the
same set of Pauli measurements), may help mitigate this potential issue.
Overall, it will be interesting to continue pushing down the general track of
time grid methods: it is certainly possible that the no free lunch theorem
somehow still holds here, but the glimmering alternative is a series of methods
with wide near-term applicability that are much more tractable than full phase
estimation.

\textbf{Acknowledgements:} RMP thanks Prof. Todd J. Mart\'inez for many
interesting discussions on classical filter diagonalization approaches and
related methods. RMP and PLM acknowledge Mr. Joseph T. Iosue and Prof. Wim van
Dam for useful discussions on the extended swap test. RMP and PLM own
stock/options in QC Ware Corporation.

\appendix

\section{Trotterized Time Propagation (Nonvariational)}

To probe the importance of the variational property in QFD, we have implemented
a nonvariational form of QFD, where formal time-translational invariance
property with exact time propagation is first used to write the QFD subspace
Hamiltonian matrix elements (and similarly the QFD subspace metric matrix
elements) as,
\begin{equation}
\mathcal{H}_{\Xi k, \Xi' k'}
=
\langle \Phi_{\Xi} | 
\hat H
e^{-i 2 \pi (k' - k) \hat H / \kappa}
| \Phi_{\Xi'} \rangle
\end{equation}
which reduces the number of formally unique matrix elements from $(2
k_{\mathrm{max}} + 1)^2$ to $4 k_{\mathrm{max}} + 1$. The Trotterization
procedure is then applied \emph{after} the formal manipulations of the matrix
elements, resulting in a physically realizable but nonvariational QFD method.

The performance of this method is drastically degraded relative to variational
QFD, to the point that nonvariational errors larger than the excitation energies
are encountered even with very small Trotter timesteps. In fact, we were only
able to achieve accurate results with the nonvariational QFD approach for
$k_{\mathrm{max}} = 1$ with overwhelmingly long Trotterizations with 100-1000
steps. This highlights the expected importance of the variational property
within the preferred QFD method developed above.

\bibliography{jrncodes.bib,refs.bib}
\bibliographystyle{aip}


\end{document}